\documentclass[aps,twocolumn,showpacs,preprintnumbers,amsmath,amssymb,a4paper]{revtex4-1}

\usepackage{graphicx}
\usepackage{dcolumn}
\usepackage{bm}
\usepackage{color}
\usepackage[T1]{fontenc}
\usepackage{textcomp}

\begin{document}
	
	\title{Effects of strain-tunable valleys on charge transport in bismuth}
	
	\author{Suguru~Hosoi$^{1,{\dag}}$}
	\author{Fumu~Tachibana$^{1}$}
	\author{Mai~Sakaguchi$^{1}$}
	\author{Kentaro~Ishida$^1$}
	\author{Masaaki~Shimozawa$^1$}
	\author{Koichi~Izawa$^1$}
	\author{Yuki~Fuseya$^2$}
	\author{Yuto~Kinoshita$^3$}
	\author{Masashi~Tokunaga$^3$}
	
	\affiliation{
		$^1$\mbox{Graduate School of Engineering Science, Osaka University, Toyonaka, Osaka, 560-8531, Japan}
		$^2$\mbox{Department of Engineering Science, University of Electro-Communications,}
		\mbox{Chofu, Tokyo, 182-8585, Japan}
		$^3$\mbox{Institute for Solid State Physics, The University of Tokyo,}
		\mbox{ Kashiwa, Chiba, 277-8581, Japan}\\
		$^{\dag}$\rm{To whom correspondence should be addressed.\\E-mail: hosoi@mp.es.osaka-u.ac.jp}
	}

	

	\begin{abstract}		
The manipulation of the valley degree of freedom can boost the technological development of novel functional devices based on valleytronics. The current mainstream platform for valleytronics is to produce a monolayer with inversion asymmetry, in which the strain-band engineering through the substrates can serve to improve the performance of valley-based devices. However, pinpointing the effective role of strain is inevitable for the precise design of the desired valley structure. Here, we demonstrate the charge transport under continuously controllable external strain for bulk bismuth crystals with three equivalent electron valleys and one hole valley. The strain response of resistance, namely elastoresistance, exhibits the evolutions in both antisymmetric and symmetric channels with decreasing temperature. The elastoresistance behaviors mainly reflect the significant changes in valley density depending on the symmetry of induced strain, evidenced by our strain-dependent quantum oscillation measurements and first-principle band calculations under strain. These facts suggest the successful tune and evaluation of the valley populations through strain-dependent charge valley transport.
	\end{abstract}

	\maketitle

\section{Introduction}
 Quantum degrees of freedom provide a central ingredient for the applications of functional electronic devices. Among them, the local conduction-band minimum, valley, is attracting attention as a key element for high-profile valleytronics, subsequently to charge for electronics and spin for spintronics\cite{SchaibleyNatRevMat2016}. A fundamental step for exploiting the valley degrees of freedom is the development of the method for lifting and monitoring the degenerated energy of valleys at different positions in momentum space. Successful valley selection has so far been demonstrated by various strategies: strain for 2D electron-gas systems in AlAs heterostructure\cite{PhysRevLett.97.186404}, electric field for diamonds\cite{IsbergNatMater2013}, polarized light for transition metal dichalcogenides\cite{ZengNatNanotech2012,MakNanotecqh2012,CaoNC2012}, and magnetic field for bismuth\cite{ZhuNatPhys2012,KulcherNatMater2014}. In addition, direct assessments of valleys have been reported in sophisticated spectroscopy measurements\cite{ZengNatNanotech2012,MakNanotecqh2012,CaoNC2012}. For the further development of potential valleytronic applications, it requires simpler methods that serve as both a controller and a barometer of valley degrees of freedom.	

  One of the practical approaches is to control and evaluate valley degrees of freedom through electrical transport. An appropriate material for this situation is a single-element semimetal bismuth with three equivalent electron valleys and one hole valley\cite{JPIssi,doi:10.7566/JPSJ.84.012001,Zhu_2018}. A strong magnetic field ($B>40$ T) can completely polarize their electron valleys depending on the direction of the applied magnetic field. For example, under a magnetic field along the binary direction, one electron valley survives, whereas the other two electron valleys disappear; this is completely opposite to the case for the field along the bisectrix\cite{ZhuNatCommun2017,IwasaSciRep2019}. Furthermore, bismuth exhibits characteristic field-angle dependent orbital magnetoresistance that can be captured by the classical transport theory with assumed ellipsoid shape of mobility tensors for one of the three equivalent electron valleys and hole valley\cite{PhysRevX.5.021022}, respectively. Therefore, bismuth is a good platform to describe the valley-dependent charge transport. However, even a few Tesla of magnetic field that is enough to induce finite valley polarization secondarily causes prominent quantum oscillations, which makes it complicated beyond the scope of this classical treatment. Alternatively, our focused strain is expected to be an effective tool to simply lift valley degeneracy\cite{BrandtSovPhys}. 
  
 The potential roles of strain in valley materials are not only limited to produce valley polarization, but expand to band engineering to acquire an ideal valley structure. As mentioned above, symmetry-breaking anisotropic strain can directly break the degeneracy of valleys. On the other hand, in-plane symmetric strain does not induce valley polarization but alternatively serves as tuning the band gap that determines the capability of potential device applications by shifting the energy level of the valleys. In fact, an enormous effort mainly using epitaxial strain has been directed to increasing the band gap for valley materials: graphene\cite{ZhouNatMat}, germanium\cite{JurgenNatPhoto}, and transition metal dichalcogenides\cite{PhysRevLett.105.136805}. Thus, utilizing each symmetry channel of strain is expected to be useful for the precise design of the valley structure, but no such systematic experiments have been conducted. Here, we demonstrate simultaneous control and evaluation of valleys in bismuth via charge valley transport under the uniaxial stress. In order to clarify the effective role of the induced strain depending on each symmetry channel, we performed symmetry-resolved elastoresistance measurements of bismuth. Since the crystal structure of bismuth has two bilayers within the hexagonal structure\cite{HOFMANN2006191}, the demonstration of strain-engineerable valleys in bismuth can provide the significant insights for tuning valley profiles of promising valleytronics candidates via epitaxial strain, such as graphene and atomically thin transition metal dichalcogenides with a hexagonal crystal structure.       
    
\section{Methods}
\begin{figure}
\centering
\includegraphics[angle=0,width=90mm]{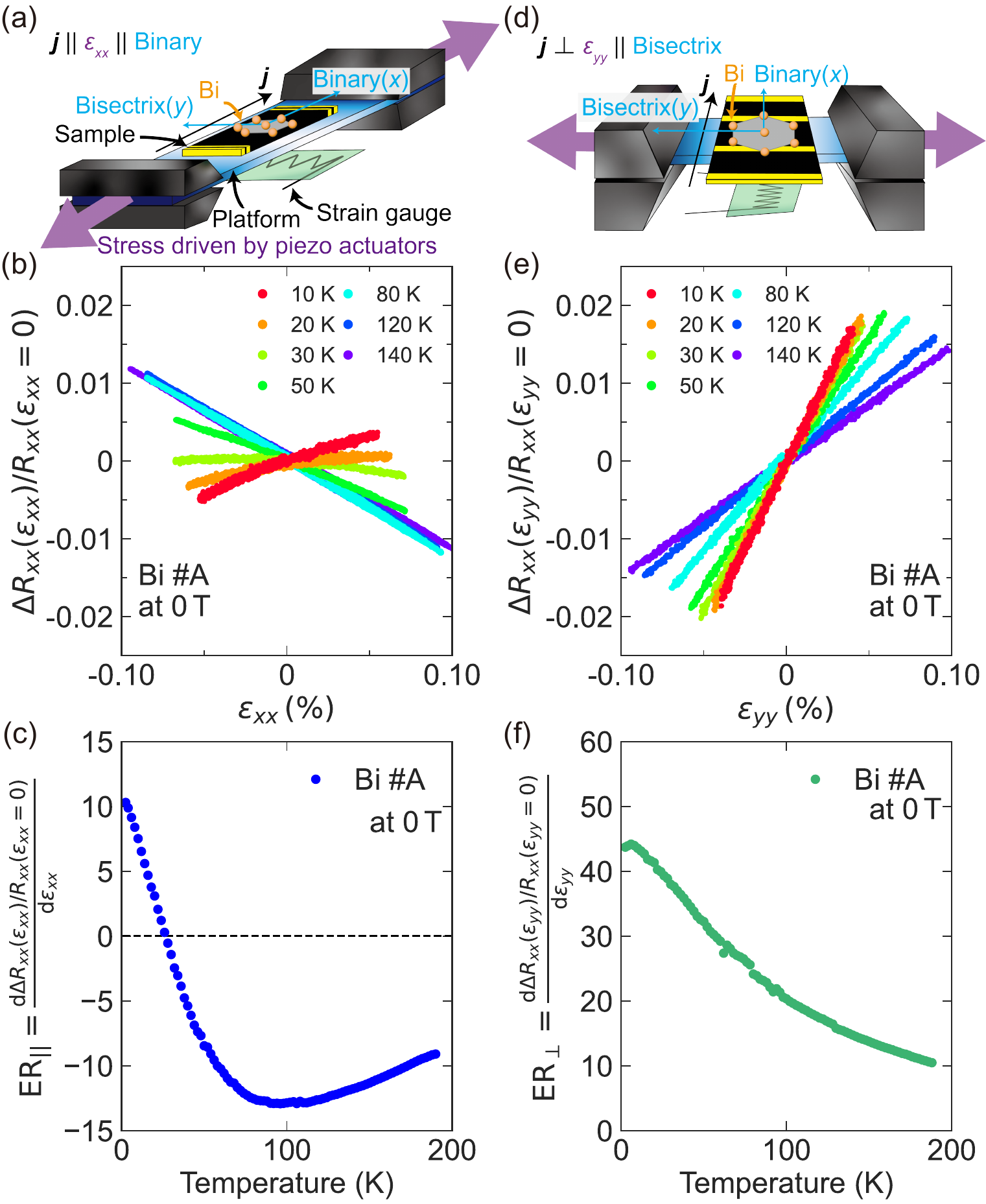}
\caption{Two experimental geometry of elastoresistance measurements in bismuth. (a-c) Strain $\varepsilon_{xx}$ dependence of resistance $R_{xx}$ along binary (b) and temperature dependence of elastoresistance ER$_{||}$ (c) in the parallel geometry (a). Sample is glued on the platform for applying strain. (d-f) Strain $\varepsilon_{yy}$ dependence of resistance (e) and temperature dependence of elastoresistance ER$_{\perp}$ (f) in the perpendicular geometry (d). \#A represents the badge number of the samples.}
\label{ER_0T}
\end{figure}

 Sample specimens were firstly spark-cut from the ingot of single-crystal bismuth grown by the Czochoralski method. Then, those specimens were cleaved and cut to achieve suitable dimensions for elastoresistance measurements: typically $\sim$1 mm (binary:$x$) $\times$ 400 {\textmu}m (bisectrix:$y$) $\times$ 60 {\textmu}m (trigonal:$z$). Uniaxial stress was applied to samples attached on the rigid platform made of titanium to achieve large strain without the destruction of the samples\cite{10.1063/5.0008829} using the home-built piezo-driven apparatus based on the design originally reported in Ref.\cite{10.1063/1.4881611}. To elucidate the symmetry-resolved strain response of bismuth, we have measured strain-induced changes in binary-direction resistance $\Delta R_{xx} (\varepsilon_{ii})= R_{xx}(\varepsilon_{ii}) - R_{xx}(\varepsilon_{ii}=0)$ ($i$ represents $x$ or $y$)  under the two different experimental geometries: the applied strain along binary $\varepsilon_{xx}$ in the longitudinal geometry (Fig.\,\ref{ER_0T}(a)) and that along bisectrix $\varepsilon_{yy}$ in the transverse geometry (Fig.\,\ref{ER_0T}(b)). The resistive strain gauge was attached to the backside of the platform to evaluate the amount of one main component of the induced strain. The strain evaluated by the strain gauge corresponds to the sample strain along binary $\varepsilon_{xx}$ in the longitudinal geometry and that along bisectrix $\varepsilon_{yy}$ in the transverse geometry, respectively. From these two experiments, we can obtain two elastoresistance:
	\begin{equation}
	{\rm ER}_{||} = \frac{{\rm d} \Delta R_{xx}(\varepsilon_{xx})/R_{xx}(\varepsilon_{xx}=0)}{{\rm d} \varepsilon_{xx}},
	\end{equation}
	
and
	\begin{equation}
	{\rm ER}_{\perp} = \frac{{\rm d} \Delta R_{xx}(\varepsilon_{yy})/R_{xx}(\varepsilon_{yy}=0)}{{\rm d} \varepsilon_{yy}}.
	\end{equation}

  Following the differential elastoresistance analysis, which was originally introduced in the iron-based superconductors\cite{PhysRevB.88.085113}, the symmetry-resolved elastoresistance can be decomposed into two parts:
  \begin{equation}
  	 {\rm ER}_{\rm sym} = \frac{1}{(1-\nu_p)}({\rm ER}_{||}+{\rm ER}_{\perp}),
  \end{equation}
  and
    \begin{equation}
    {\rm ER}_{\rm anti} =  \frac{1}{(1+\nu_p)}({\rm ER}_{||}-{\rm ER}_{\perp}),
  \end{equation}
  where $\nu_p$ is an effective Poisson ratio of the platform directly measured by the strain gauges($\nu_p \sim 0.197$). ER$_{\rm sym}$ and ER$_{\rm anti}$ represent elastoresistance response against the isotropic symmetric strain $\varepsilon_{\rm sym}=(\varepsilon_{xx}+\varepsilon_{yy})/2$ and the anisotropic antisymmetric strain $\varepsilon_{\rm anti} = (\varepsilon_{xx}-\varepsilon_{yy})/2$, respectively. For simplicity, we consider strain components only in the $xy$ plane.
 
  For exploring the relationships between strain-controlled valley density and elastoresistance results, we developed the minimum classical framework that provides the phenomenological understanding of how the strain-modified valley population affects charge transport. In this model, we assume the rigid band approximation that conductivity under strain only changes its carrier density term. This model provides the quantitative evaluation of the valley susceptibility $\chi_{\rm \Gamma}$, which describe the controllability of the valley density against each symmetry channel $\Gamma$ of the applied strain: symmetric strain $\varepsilon_{\rm sym}$ and antisymmetric strain $\varepsilon_{\rm anti}$. To clarify the expected strain-modified valley density from these obtained valley susceptibilities both experimentally and theoretically, we have performed quantum oscillation measurements under strain and density functional theory (DFT) calculations. Both results are successfully explained by the valley susceptibilities evaluated by the elastoresistance measurements.
  
 DFT calculations were performed using BAND software of Amsterdam Modeling Suite \cite{PhysRevB.44.7888,BAND}. We employed the Generalized Gradient Approximation (GGA) with the Perdew-Burke-Ernzerhof exchange-correlation functional and triple-zeta-polarized basis sets. The relativistic effects were considered by the noncolinear method. It is well known that the GGA of DFT overestimates the direct gap at the L-points \cite{Liu1995,PhysRevB.91.125129}. However, it has also been shown that the GGA yields a result qualitatively consistent with the more accurate approximation, such as the quasiparticle self-consistent $GW$ calculation \cite{PhysRevB.91.125129}. Therefore, the present results should qualitatively capture the band-structure change against the strain.
      We modified the $x$-, and $y$-components of the basic translation vector $\bm{a}=(a_x, a_y, a_z)$ as $a'_x = a_x (1 + \varepsilon_{xx})$ and $a'_y = a_y (1+\varepsilon_{yy})$, keeping $a_z$ unchanged. For the symmetric strain, we set $\varepsilon_{xx} = \varepsilon_{yy}$, while we set $\varepsilon_{xx} = -\varepsilon_{yy}$ for the antisymmetric strain.
      
Quantum oscillation measurements were performed under the in-situ set-up for elastoresistance measurements in order to directly evaluate valley density. To align the magnetic field along binary direction in both two experimental geometries, we used two superconducting magnets: vertical magnetic field up to 7 T by solenoid magnet for the longitudinal geometry and the horizontal field up to 4 T by split magnet for the transverse geometry. Although the available magnetic field window is limited, two of the three electron pockets reaches the quantum limit(QL) at 1.5 T under magnetic field along binary owing to the smallness of the Fermi energy in bismuth, which enable us to study the strain-dependence of valley density. We evaluated the valley susceptibilities based on the strain-dependent QLs apart from the elastoresistance analysis.
  	
\section{Results}
\subsection{Elastoresistance measurements}
Response in resistance of bismuth against the applied strain is summarized in Fig.$\,$\ref{ER_0T}. Elastoresistance of bismuth exhibits contrasting results between two experimental geometries. First, we measured longitudinal elastoresistance ER$_{\rm ||}$ in the parallel geometry depicted in Fig.$\,$\ref{ER_0T}(a), where the applied current $j$ and induced strain $\varepsilon_{xx}$ are along the binary ($x$) direction as $j \,||\, \varepsilon_{xx} \,|| \,{\rm binary}$. ER$_{\rm ||}$ changes its sign from negative to positive on cooling with a broad minimum structure, as shown in Figs.$\,$\ref{ER_0T}(b),(c). After the longitudinal geometry experiment, we then measured transverse one ER$_{\perp}$ in the perpendicular geometry $j \,\perp\, \varepsilon_{yy} \,|| \,{\rm bisectirx}$ (Fig.$\,$\ref{ER_0T}(d)). In contrast to ER$_{\rm ||}$, ER$_{\perp}$ monotonically increases with decreasing temperature, as shown in Figs.$\,$\ref{ER_0T}(e),(f). The observed strain directional difference is consistent with the previous comparable study above liquid nitrogen temperature using thin film samples, including each sign and amplitude\cite{Koike_1966,RevPhysAppl1979}, although we note there are still other previous studies that measured other elastoresistance coefficients in  different experimental geometry\cite{PhysRev.135.A708,PhysRev.149.485}. The essential differences in temperature dependence between ER$_{\rm ||}$ and ER$_{\rm \perp}$ may reflect the mixing contributions from two symmetry channels.

\begin{figure}[t]
\centering
\includegraphics[angle=0,width=85mm]{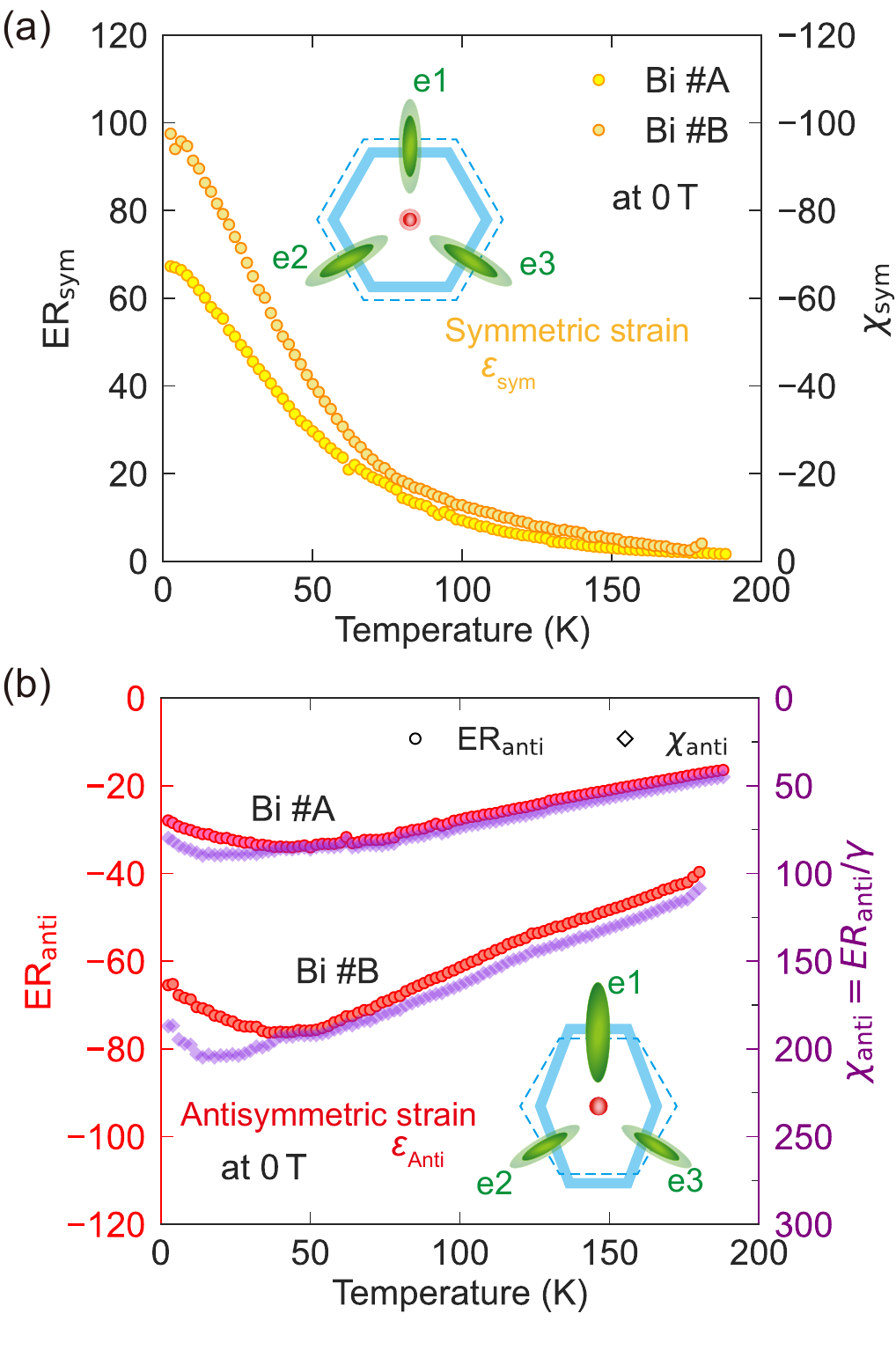}
\caption{Symmetry-resolved elastoresistance response of bismuth. (a) Symmetric components of elastoresistance for left axis. Right axis represents the symmetric valley susceptibility based on the relation ER$_{\rm sym} = - \chi_{\rm sym} $. To ensure the reproducibility, we shows the results of two samples, \#A and \#B. Inset shows a schematic picture of changes in valley structure induced by tensile symmetric strain.  Blue dashed and solid lines depict original and strain-decreased Brillouin zones (BZs), respectively. We note that the change in the shape of BZ is opposite to the strain in the real space. Green ellipsoids and red circles represents electron and hole valleys, respectively. (b) Antisymmetric components of elastoresistance represented by pink circles for left axis. Purple diamonds represent valley susceptibility evaluated by $\chi_{\rm anti}= {\rm ER}_{\rm anti}/\gamma(T)$, whose scale is shown in the right axis. Inset represents schematics of valley polarizations induced by antisymmetric strain. Blue dashed and solid lines depict original and strain-deformed BZs, respectively.}
\label{ERsym_0T}
\end{figure}

To elucidate the origin of these strain direction-dependent behaviors, we resolved two components of elastoresistance by combing the results of both experimental geometries: symmetric component ER$_{\rm sym}$ and antisymmetric component ER$_{\rm anti}$, as shown in Figs.$\,$\ref{ERsym_0T}(a),(b). We have performed several samples to ensure the reproducibility of the results(see also Appendix A1). In addition, our resolved ER$_{\rm sym}$ well agrees with relevant elastoresistance coefficients directly measured in the early study\cite{PhysRev.135.A708}(See Appendix A1), supporting the validity of our symmetry-resolved analysis. In high temperature regions, ER$_{\rm anti}$ dominates over ER$_{\rm sym}$. The magnitude of both ER$_{\rm anti}$ and ER$_{\rm sym}$ exhibits the enhancement on cooling with opposite signs, indicating the sensitivity of bismuth against multiple symmetry channel of the strain. However, this ER$_{\rm anti}$ is saturated roughly around $T \sim 50 \, {\rm K}$, while ER$_{\rm sym}$ shows continuous enhancements.  This symmetry crossover from antisymmetric to symmetric response reflects a broad minimum in ER$_{\rm ||}$ with sign change, as shown in Fig.$\,$\ref{ER_0T}(c). By contrast, in the perpendicular geometry, both two channels of elastoresistance give cooperative contributions, leading to the strong enhancements of ER$_{\perp}$, as shown in Fig.$\,$\ref{ER_0T}(f).

\subsection{Valley susceptibility analysis}
We now address the microscopic mechanism behind this strain-sensitive charge transport. The previous magnetostriction results provide valuable insights into the intimate relationships between strain and valley density\cite{PRSL_Kapitza,PRSL_Shoenberg,PhysRevLett.11.331,JPMichenaud_1981,PhysRevB.26.2552,KulcherNatMater2014}. Very large magnetostriction observed in bismuth can be attributed to field-induced changes in valley density, which can be enhanced by carrier-transfer process between multivalleys\cite{PhysRevB.26.2552}. This fact leads us to expect its reverse case: the external strain can cause significant changes in valley density. To elucidate the pure effect of strain-induced changes in valley density on transport, we start from the classical framework introduced in Ref.\cite{PhysRevX.5.021022}, where the conductivity is described by the summation of each valley contribution. Once the mobility tensor of one electron valley and hole valley is fixed, this basic and simple framework successfully captures transport properties of bismuth\cite{PhysRev.181.1070,J_PMichenaud_1972}, including the even more puzzling field-angle dependence of magnetoresistance\cite{PhysRevX.5.021022}. However, when we extend the scope of this model to transport under strain, the form of conductivity under strain become generally complicated because the application of strain alters both carrier density and mobility. Here, for simplicity, we propose the carrier-based model under strain $\varepsilon$ for each valley with only the strain-induced changes in carrier density $\Delta n_i (\varepsilon)= n_i(\varepsilon)- n_{i}(\varepsilon=0)$, where $n_i$ represents the carrier density of each valley with index $i$. In fact, the importance of this carrier density term for describing elastoresistance behaviors has also been acknowledged for the WTe$_2$\cite{doi:10.1073/pnas.1910695116}, which is one of the well-known semimetals with a small carrier concentration, just like bismuth. We also add the symmetry-dependent changes in valley density to reflect the valley degree of freedom of bismuth, which cannot be explored in WTe$_2$ due to the low crystal symmetry and the lack of valley degrees of freedom. Here, to decribe $\Delta n_i (\varepsilon)$, we introduce symmetry-decomposed strain-valley susceptibility $\chi^i_{\Gamma} = (1/n_i (\varepsilon=0) ){\rm d}n_i/{\rm d} \varepsilon_{\Gamma}$. As described below, this simplified model can essentially capture the elastoresistance of bismuth.

\begin{figure*}
	\centering
	\includegraphics[angle=0,width=150mm]{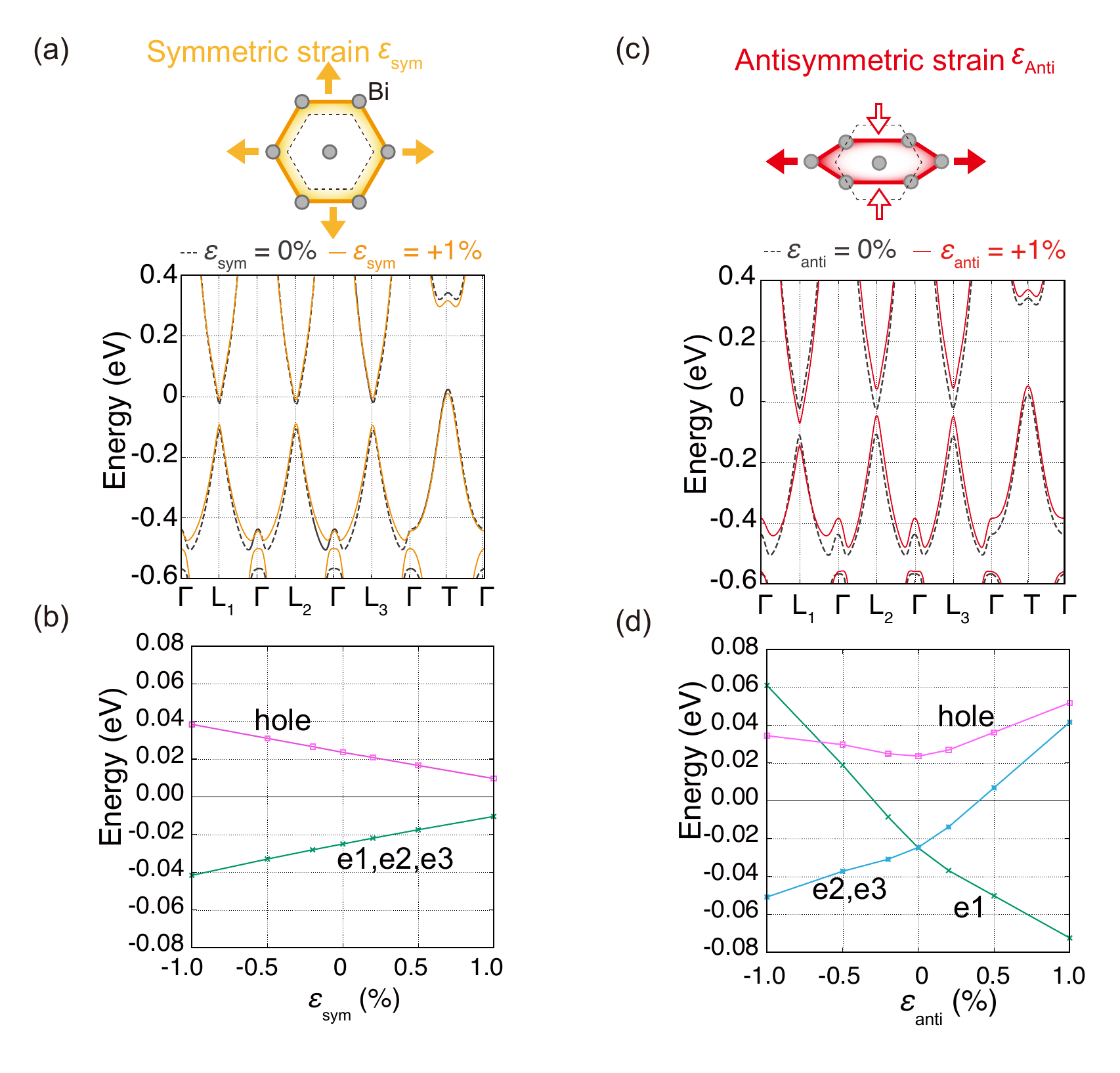}
  \vspace{-10mm}
	\caption{DFT band calculation results under strain. (a),(c) (upper panel) The schematic images of the applied strain-induced changes in the trigonal plane crystal lattice: symmetric strain (a) and antisymmetric strain (c). (lower panel) Band structure near the the electron (L$_1$, L$_2$, L$_3$) and hole (T) pockets under symmetric strain $\varepsilon_{xx} = \varepsilon_{yy}=0.01$ (a) and antisymmetric strain $\varepsilon_{xx} = -\varepsilon_{yy}=0.01$ (c).The dashed lines indicate the band structures for zero strain $\varepsilon_{xx} = \varepsilon_{yy}=0$.(b),(d) Energy of the bottom of the electron and the top of the hole bands as a function of symmetric (b) and  antisymmetric strains (d).}
	\label{DFT}
\end{figure*}

The modification of valley structures is constrained by the symmetry of the lattice deformation. Isotropic symmetric strain $\varepsilon_{\rm sym}$ preserves the rotational symmetry underlying the crystal lattice, leading to the uniform change of valley population without breaking the equivalence of three electron valleys. Adding the charge neutrality condition, the valley population varies with symmetric strain as
\begin{equation}
  \Delta n_{\rm e1} = \Delta  n_{\rm e2} = \Delta n_{\rm e3} = \Delta n_{\rm hole}/3 = n \chi_{\rm sym}\varepsilon_{\rm sym},	
\end{equation}
 where $n$ represents the valley density for one electron valley at ambient stress. In this situation, $\varepsilon_{\rm sym}$ only changes the total carrier number described as $\chi_{\rm sym}$, and hence straightforwardly connects with elastoresistance within the carrier-based model neglecting the strain-induced modification of mobility(see Appendix A2) as below: 
 \begin{equation}
 	 {\rm ER}_{\rm sym} = - \chi_{\rm sym}.
 \end{equation}
Based on this model, the observed positive sign of the symmetric elastoresistance ER$_{\rm sym}$ indicates that a tensile strain, which reduces the size of the Brillouin zone, decreases the carrier density of each valley, as shown in the inset of Fig.$\,$\ref{ERsym_0T}(a). This behavior is consistent with the previous first principle study reporting that the overlap of the in-direct band gap between the hole and electron becomes small by the expansion of the trigonal plane crystal lattice\cite{PhysRevB.91.125129} and our calculations discussed below (see Figs.$\,$\ref{DFT}(a),(b)). This band modification reflects the charge neutrality of semimetal and becomes significant particularly at low temperatures, where only low energy bands near the Fermi level become relevant. In fact, the elastoresistance of WTe$_2$ at low temperatures is also attributed to the charge neutral band modification\cite{doi:10.1073/pnas.1910695116}. Thus, ER$_{\rm sym}$ clearly visualizes the temperature evolutions of uniform energy shifts of electron and hole valleys, as shown in Fig.$\,$\ref{ERsym_0T}(a).

On the other hand, symmetry-breaking antisymmetric strain $\varepsilon_{\rm anti}$ can make a difference in the valley polarization in one valley e1 and the other two valleys e2/e3:
\begin{eqnarray}
	  \Delta n_{\rm e1} &=& n \chi_{\rm anti} \varepsilon_{\rm anti}, \\
	  \Delta n_{e2/e3} &=& -n\chi_{\rm anti} \varepsilon_{\rm anti}/2 .
\end{eqnarray}
 Now, $\chi_{\rm anti}$ represents the valley susceptibility that evaluates the sensitivity of the valley polarization against applied symmetry-breaking antisymmetric strain, which corresponds to so-called nematic susceptibility applied to various iron-based superconductors\cite{doi:10.1126/science.1221713,doi:10.1126/science.aab0103,doi:10.1073/pnas.1605806113,doi:10.1073/pnas.2110501119}. In contrast to the case of $\chi_{\rm sym}$, the relationships between $\chi_{\rm anti}$ and ER$_{\rm anti}$ depend on the anisotropy of valley mobility $\gamma$:
\begin{equation}
	{\rm ER}_{\rm anti} = \gamma \chi_{\rm anti}.
\end{equation}
The detailed derivation of this relation is provided in the Appendix A2. The relevant anisotropic factor in this experimental geometry is evaluated as $\gamma \sim -0.35$ at low temperatures based on the previous studies\cite{PhysRevX.5.021022,PhysRev.181.1070}. The negative sign of $\gamma$ comes from the fact that valley e1 has higher mobility along binary than the other electron valleys e2/e3, leading to the conductivity improvements by positive antisymmetric strain-induced increases of the e1 valley density. Figure$\,$\ref{ERsym_0T}(b) depicts the overall temperature dependence of $\chi_{\rm anti}$, which incorporates the temperature dependence of $\gamma$\cite{PhysRevX.5.021022,PhysRev.181.1070,J_PMichenaud_1972} (see also Appendix A2). The estimated $\chi_{\rm anti}$ is comparable to or even larger than $\chi_{\rm sym}$, suggesting the strain sensitivity in the valley densities against both symmetric and antisymmetric strains.

\begin{figure*}
  \centering
  \includegraphics[angle=0,width=150mm]{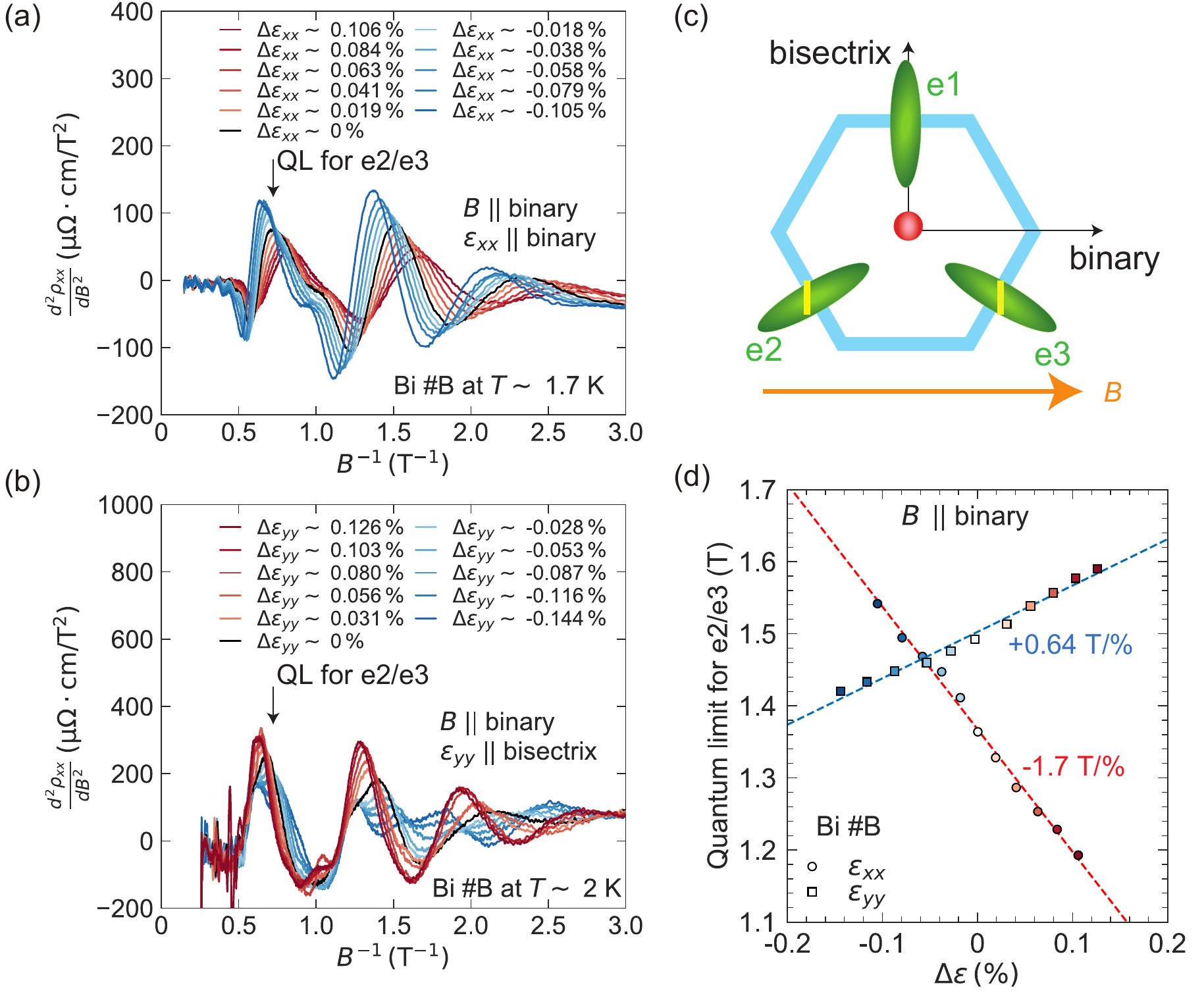}
  \vspace{-5mm}
  \caption{Strain-dependent quantum oscillations of sample \#B along $B$ $||$ binary under two experimental geometries. (a),(b) Shubnikov-de Haas oscillations under the parallel (a) and perpendicular geometries (b). The black arrow indicates quantum limit (QL) for electron valleys e2/e3 at zero strain point. (c) Schematic illustration of the Fermi surface area originating from the quantum oscillations indicated by yellow lines on the valleys e2/e3. (d) Both $\varepsilon_{xx}$ and $\varepsilon_{yy}$ dependent shifts of QLs of electron valleys e2/e3.}
  \label{QO_strain_dep.pdf}
  \end{figure*}

\subsection{DFT calculation under strain}
 These strain-modified valley structures are successfully visualized by our density functional theory (DFT) calculations under strain, as shown in Fig.$\,$\ref{DFT}. Three electron valleys e1, e2, e3 and hole valley are located at three equivalent L-points labeled as  L$_1$, L$_2$, L$_3$ and T-point, respectively. Under a tensile symmetric strain, all electron valleys equivalently shift upward while hole valley does downward as shown in Fig.$\,$\ref{DFT}(a), leading to the uniform reduction of each valley density. As shown in Fig.$\,$\ref{DFT}(b), the symmetric strain linearly shifts each valley within the calculated region $\varepsilon_{\rm sym} \pm 1 \%$, justifying the strain-linear response analysis of valley density. On the other hand, the antisymmetric strain induces electron valley polarization as expected from $\chi_{\rm anti}$: for instance, the positive antisymmetric strain increases e1 valley density but decreases e2/e3 valley densities, as shown in Fig.$\,$\ref{DFT}(c). The detailed strain dependence of valleys shown in Fig.$\,$\ref{DFT}(d) clearly depicts the switch of the valley polarization across $\varepsilon_{\rm anti} = 0$. The carrier number of isotropic hole valley should change equally under $\pm \varepsilon_{\rm anti}$, but anisotropic shifts of electron valleys result in asymmetric even-functional strain dependence of hole valley through the charge neutrality of semimetals. The hole valley changes little under the perturbative small strain used in the elastoresistance study, implying the transport under $\varepsilon_{\rm anti}$ dominated by electron valleys. These strain-induced changes in valley density are caused by the simple energy shifts of the valleys with keeping their band shapes. This fact suggests the validity of the rigid band approximation; thus, charge transport under strain should be dominated by strain-induced changes in valley density since the fermi velocity is nearly unchanged. These DFT calculation results are essentially consistent with the changes in valley density elucidated by the elastoresistance signal based on the simple carrier-based classical transport analysis, supporting the successful strain-tuning of valleys and evaluation of its effect on transport.

\subsection{Quantum oscillation measurements under strain}
  To further strengthen our discussions, we try to directly evaluate the strain-induced changes in valley populations through strain-dependent quantum oscillations of bismuth at the lowest temperature measured. A magnetic field is applied along the binary direction in both strain geometries, as shown in Figs.$\,$\ref{QO_strain_dep.pdf}(a)(b). Three clear peaks are observed in the second field derivative of resistivity derived from Shubnikov-de Haas(SdH) oscillations of electron valleys e2/e3, determined by the well-defined Landau spectrum of bismuth\cite{Zhu_2018,PhysRevB.84.115137}(See also Fig.$\,$\ref{QO_strain_dep.pdf}(c)). These peaks exhibit strain sensitivity; for the parallel geometry, positive strain $\varepsilon_{xx}$ shifts the peak position to a lower field side, as shown in Fig.$\,$\ref{QO_strain_dep.pdf}(a), evidencing the shrink of the electron valleys e2/e3; by contrast, each peak exhibits shifts toward higher field region under positive strain $\varepsilon_{yy}$ for the perpendicular geometry, as shown in Fig.$\,$\ref{QO_strain_dep.pdf}(b). Here, due to the small number of observable peaks, we focus on the QLs to estimate the strain dependence of the valley density instead of using the conventional fast Fourier transformation (FFT) analysis. Both strain $\varepsilon_{xx}$ and $\varepsilon_{yy}$ dependent QLs for valleys e2/e3 are shown in Fig.$\,$\ref{QO_strain_dep.pdf}(d). Combing these two results of the strain-controlled QLs gives another evaluation of valley susceptibility: $\chi_{\rm sym}^{\rm QO} \sim -100$ and $\chi_{\rm anti}^{\rm QO} \sim 280$(see Appendix A3). These values are qualitatively consistent with the evaluations by elastoresistance around the same low temperatures, including each magnitude and sign (see Figs.$\,$\ref{ERsym_0T}(a)(c)).

\section{Discussion}
As described above, quantum oscillation measurements demonstrate that strain-induced valley density change coincides with valley susceptibility described by elastoresistance, supporting the validity of our proposed simple carrier-based transport model under strain. Furthermore, it is worth noting that the evaluation of $\chi$ also agrees with our analysis based on the deformation potentials \cite{PhysRev.174.782} within the framework introduced in the early magetostriction study\cite{PhysRevB.26.2552}(see Appendix A4). These facts strongly suggest that the sensitivity of valley density against strain is very high enough for bismuth to justify the simple phenomenological treatments between valley density and strain at least at low temperatures. Returning to our original motivation, antisymmetric strain successfully tunes electron valley degeneracy, which can be evaluated by $\chi_{\rm anti}$. In addition, $\chi_{\rm anti}$ develops with cooling, suggesting the manipulation capability of valley degrees of freedom especially at low temperatures. Furthermore, large $\chi_{\rm sym}$ suggests that a tensile symmetric strain efficiently suppresses the indirect gap of bismuth, which contributes to enhance another aspect of valley capability.
  
  Finally, we discuss the possibility of the nematic aspects of bismuth with three equivalent electron valleys. The nematic state of bismuth are described as valley polarized states, which can be classified into the novel $Z_{3}$ nematicity recently discussed in various materials such as magnetism\cite{LittleArielleNatMater2020}, charge density wave\cite{LinpengNature2022}, and nematic superconductivity\cite{ChangwooNC2020}. In fact, the possibility of valley nematic states in bismuth has been discussed in low-temperature regions under magnetic field in both bulk \cite{ZhuNatPhys2012,PhysRevX.5.021022} and surface states\cite{doi:10.1126/science.aag1715}, although the former results are recently attributed to the extrinsic effects due to the boundary conductance\cite{KangNC2022}. In that sense, the direct evaluation of valley density in the present study may demonstrate the effective role of strain in controlling these $Z_{3}$ orders.  Increasing $\chi_{\rm anti}$ at low temperatures seems not to deny the putative nematic state in bismuth. In fact, iron-based superconductors are the representative metals that exhibit such a large ER$_{\rm anti}$ comparable to bismuth, owing to the critical divergence of nematic susceptibilities\cite{doi:10.1126/science.1221713,doi:10.1126/science.aab0103,doi:10.1073/pnas.1605806113,doi:10.1073/pnas.2110501119}. However, such an enhancement in iron-based superconductors generally occurs in only one symmetry channel since ordinary nematic materials are sensitive to only the specific direction of strain that couples with the symmetry of their own nematicity. Therefore, the evolution of $\chi_{\rm sym}$, which reaches a comparable magnitude to $\chi_{\rm anti}$, clearly demarcates bismuth from simple nematic materials. The observed $\chi_{\rm anti}$ of bismuth does not necessarily pinpoint rotational symmetry breaking field. Rather, simultaneous large $\chi_{\rm sym}$ and $\chi_{\rm anti}$  describe the sensitivity to any external perturbative stress field; this strain sensitivity over multiple symmetry channels possibly originates from the nature of semimetal with the smallness of fermi energy and charge neutrality. The mixing ratio of induced symmetric strain to antisymmetric strain by applied uniaxial pressure can change through the aspect ratio of the crystal, and as a result the fabrication of sample dimension can be one tool to tune the desirable valley profiles.

	\section*{Acknowledgements}
We thank M. Hecker and J. Scmalian for helpful discussions. We also thank N. Miura for the provided bismuth samples. S.H. thanks J. Bartlett, A. Steppke, and C. W. Hicks for making him aware of the importance of studying strain-response of bismuth and sharing the the technique for developing the strain cell through the other collaboration works. This work was supported by Grants-in-Aid for Scientific Research (KAKENHI) (Nos. JP18H01167, JP20K20901, JP22H01939, JP22K03522, JP22K18690, JP23H00268, JP23H04862, JP23K17879) and Grand-in-Aid for Scientific Research on innovative areas ``Quantum Liquid Crystals'' (No. JP20H05162) from Japan Society for the Promotion of Science. S.H. and M. Shimozawa were supported by the Multidisciplinary Research Laboratory System (MRL), Osaka University, respectively.


 \section*{APPENDIX}
 \subsection*{1. Reproducibility of elastoresistance\label{Sec:APP}}
  \begin{figure}
    \centering
    \includegraphics[angle=0,width=85mm]{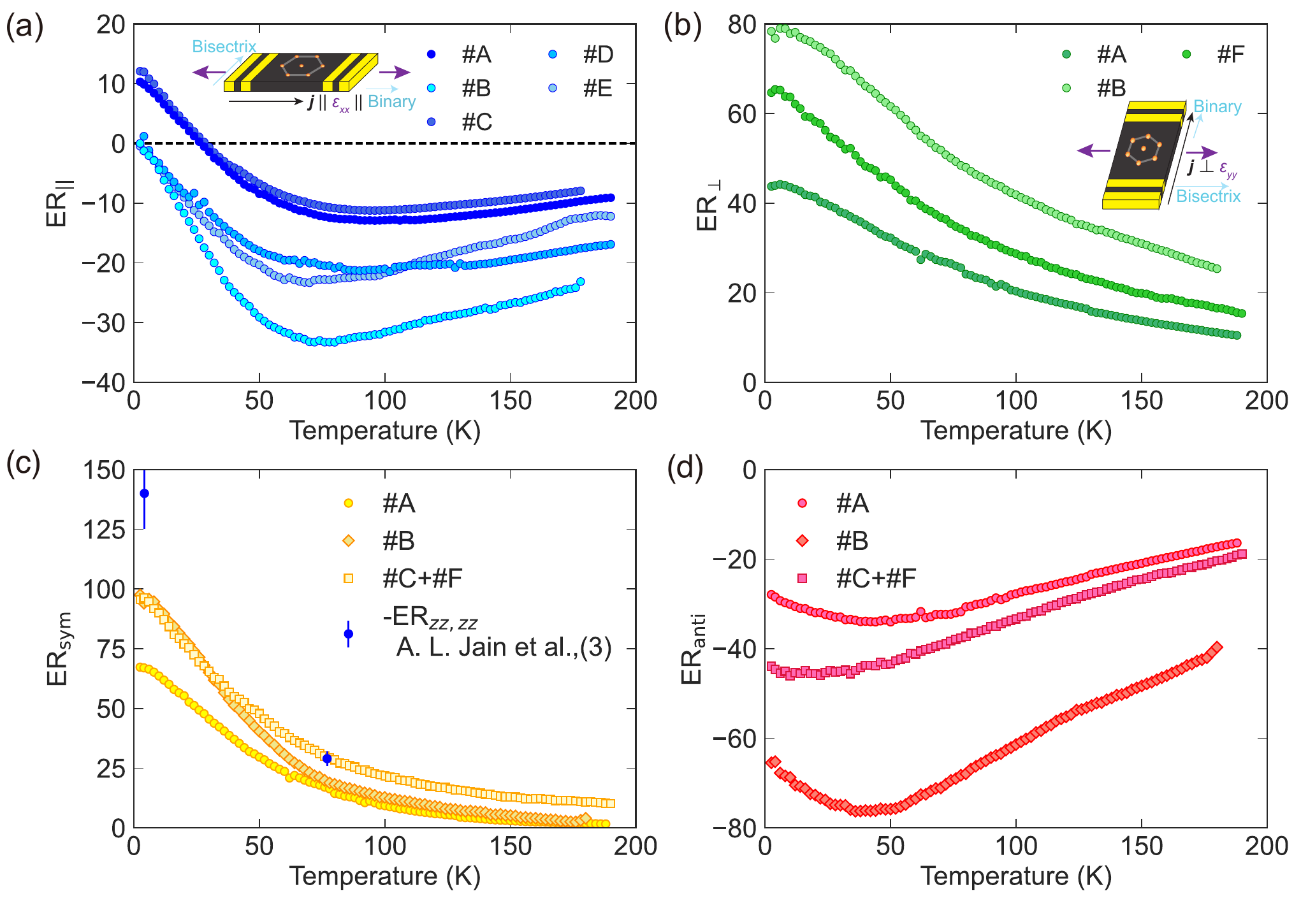}
    \caption{Two experimental geometry of elastoresistance measurements (A,B) and symmetry-decomposed elastoresistance (C,D) in bismuth for several samples. As for a reference, another type of symmetric elastoresistance $-$ER$_{zz,zz}$ \cite{PhysRev.135.A708} is also plotted in (C).}
    \label{SM_ER_repro_check}
    \end{figure}
 
One of the difficulties in quantitative analysis is that practically induced strain strongly depends on the experimental conditions; for example, sample dimension greatly affects the strain transmission rate\cite{doi:10.1126/science.aab0103,10.1063/5.0008829}. So, we have measured several samples to check the reproducibility. We measured five samples for the parallel geometry $j || \varepsilon_{xx} || {\rm binary}$ (Fig.$\,$\ref{SM_ER_repro_check}(a)) and three samples for the perpendicular geometry $j \perp \varepsilon_{yy} || {\rm bisectrix}$ (Fig.$\,$\ref{SM_ER_repro_check}(b)). For samples \#A and \#B, we first measured ER$_{||}$. Next, we took the samples \#A and \#B off the platform and re-glued them on the platform for the ER$_{\perp}$ measurements. There are some quantitative variations in elastoresistance, but all measured samples qualitatively reproduce the essential properties; ER$_{||}$ shows crossover behaviors with a broad minimum around 80 K, whereas ER$_{\perp}$ exhibits monotonic enhancements on cooling. The same warranty can be provided for symmetry-decomposed elastoresistance, as shown in Figs.$\,$\ref{SM_ER_repro_check}(c)(d). 

 In this study, we resolved the symmteric and antisymmetric responses in elastoresistance by combing the results from two different experimental geometries. The antisymmetric component cannot be directly measurable unless this combined analysis is performed, but the symmetric component can be direcly evaluated by the only one experimental geometry, where the changes in resistivity along the trigonal ($z$) direction $(\Delta \rho / \rho)_{zz}$ are measured under the longitudinal strain along the trigonal direction $\varepsilon_{zz}$. Since in-plane symmetric strain $\varepsilon_{\rm sym} = \frac{1}{2} (\varepsilon_{xx} + \varepsilon_{yy})$ is compatible with out-of-plane strain with the opposite sign $- \varepsilon_{zz}$, this elastoresistance response ER$_{zz,zz} = \frac{{\rm d}(\Delta \rho / \rho)_{zz}}{{\rm d} \varepsilon_{zz}}$ is the essentially same with our resolved symmetric elastoresistance $-$ER$_{\rm sym}$. For comparision, the results of ER$_{zz,zz}$ reported by the ealy study\cite{PhysRev.135.A708} are also plotted in Fig.$\,$\ref{SM_ER_repro_check}(c). Both ER$_{\rm sym}$ and -ER$_{zz,zz}$ exhibit the essentially same enhancement with decreasing temperatures, suggesting the validity of our symmetry-resolved analysis.

\subsection*{2. Simple carrier model for transport under strain}

 The transport model for the valley material bismuth has been established in the previous study describing the field angle dependence of magnetoresistance\cite{PhysRevX.5.021022,JEAubrey_1971}. The essential point of this theory is introducing mobility tensors for each ellipsoidal valley of bismuth. For the electron valley e1, the mobility tensor is given as follows:
 \[
 \hat{\mu}_{\rm e1} =
 \begin{bmatrix}
\mu_1 & 0 &0 \\
0 & \mu_2 & \mu_4 \\
0 & \mu_4 & \mu_3 \\
\end{bmatrix}.
 \]
  Off-diagonal component $\mu_4$ comes from the slight tilts of the electron valley in the trigonal direction. We refer each tensor component as $\mu_{ij}^{\rm e1} (i,j = x,y,z)$: for example, $\mu_{xx}^{\rm e1} = \mu_{1}$. Threefold rotational symmetry of bismuth gives equivalence among each electron valley under $2 \pi /3$ rotation. Once the rotation matrix $\hat{R}_{\theta}$ for a rotation around the trigonal axis is introduced, the mobility tensors for the other two electron valleys can be expressed as
 \[
  \hat{\mu}_{e2} = \hat{R}_{2\pi/3}^{-1} \cdot \hat{\mu}_{\rm e1} \cdot \hat{R}_{2\pi/3},
 \]
 
  \[
  \hat{\mu}_{e3} = \hat{R}_{-2\pi/3}^{-1} \cdot \hat{\mu}_{\rm e1} \cdot \hat{R}_{-2\pi/3}.
 \]
On the other hand, the hole valley mobility tensor is given as:
\[
 \hat{\nu}_{\rm h} = 
  \begin{bmatrix}
\nu_1 & 0 &0 \\
0 & \nu_1 & 0 \\
0 & 0 & \nu_3 \\
\end{bmatrix}.
\]
Since hole valley has an ellipsoidal shape with the major axis precisely along the trigonal direction, there are no off-diagonal components, in contrast to electron valleys.
By using these mobility tensors, the conductivity of bismuth is formalized as
\[
\hat{\sigma} =  \sum_{i=1,2,3}  n_{{\rm e}i} e \hat{\mu}_{{\rm e}i} + n_{\rm h} e\hat{\nu}_{\rm h}.
\]
Here, $e$ represents elementary charge. Threefold rotational symmetry guarantees the equivalence among three electron valleys as $n = n_{\rm e1} = n_{\rm e2} = n_{\rm e3}$. In addition, the charge neutrality condition of semimetal set the constraints on the number of electrons and holes as $3 n = n_{\rm h}$. In Ref.\cite{PhysRevX.5.021022}, a magnetic field tensor is incorporated as a magnetic field effect to describe magnetoresistance in accordance with Aubrey's work\cite{JEAubrey_1971}. In contrast to this, the strain effect forcused on here is generally introduced as the change in both carrier numbers and mobility for each valley. For the case of electron valley e1, the conductivity tensor component $\sigma_{xx}^{\rm e1}$ under strain is the following first-order approximation:
 \begin{eqnarray*}
 \sigma_{xx}^{\rm e1}(\varepsilon) =&  n_{{\rm e}1} (\varepsilon=0)\,e\, \mu^{{\rm e}1}_{xx}(\varepsilon=0) \\
  &\times (1 + \frac{1}{\mu^{{\rm e}1}_{xx}(\varepsilon=0)}\frac{{\rm d} \mu^{{\rm e}1}_{xx} }{{\rm d} \varepsilon} \varepsilon + \frac{1}{n_{\rm e1}(\varepsilon=0)} 
 \frac{{\rm d}n_{{\rm e}1}}{{\rm d} \varepsilon} \varepsilon).
 \end{eqnarray*}
 
Our DFT results reveal that strain shifts only the valleys with keeping their band shapes. This fact suggest that the mobility of the each valley changes little under strain, supporting this rigid band approximation. Therefore, we adopted the carrier-based model without the strain-induced mobility changes, namely the rigid band approximation against the applied strain.
 
 Strain-induced changes in charge carrier number are described by introducing the valley susceptibility defined as $\chi = \frac{1}{n_{\rm e1}(\varepsilon=0)}\frac{{\rm d} n_{{\rm e}1}}{{\rm d} \varepsilon}$.  The conductivity tensor under strain and valley susceptibility for other valleys can be expressed in the same manner. As discussed in the main text, strain responses of carrier density change depending on the symmetry of the introduced strain. Therefore, two kinds of valley susceptibility can be introduced: symmetric valley susceptibility $\chi_{\rm sym}$ and antisymmetric valley susceptibility $\chi_{\rm anti}$. In the following, we discuss the relationships between elastoresistance and each valley susceptibility.
     
    Symmetric strain $\varepsilon_{\rm sym} = \frac{1}{2}(\varepsilon_{xx}+ \varepsilon_{yy})$ preserves the symmetry underlying lattice and uniformly changes three electron valleys: $\Delta n_{{\rm e}1} =\Delta n_{{\rm e}2} = \Delta n_{{\rm e}3} =n \chi_{\rm sym} \varepsilon_{\rm sym} $, where $\Delta n_{{\rm e}i}$ represents strain-induced changes in each valley density as $\Delta n_{{\rm e}i}= n_{{\rm e}i}(\varepsilon) - n$ and we introduce common valley density $n$ among electron valleys at ambient stress.
    Charge neutrality conditions constrain the changes in hole valley density as $\Delta  n_{\rm h} = \sum_{i = 1,2,3} \Delta n_{{\rm e} i} = 3 n \chi_{\rm sym} \varepsilon_{\rm sym}$. Using these modifications of carrier density, the conductivity tensor under symmetric strain is given as:    
  
    \[
    \hat{\sigma} (\varepsilon_{\rm sym})= n e (1 + \chi_{\rm sym} \varepsilon_{\rm sym})(\hat{\mu}_{{\rm e}1} + \hat{\mu}_{{\rm e}2} + \hat{\mu}_{{\rm e}3}+ 3\hat{\nu}).
    \]
    By using this equation, the strain-induced changes in the binary-direction resistivity $\rho_{xx}$ is given as:
 \begin{eqnarray*}
    &&\Delta \rho_{xx} (\varepsilon_{\rm sym})/\rho_{xx}(\varepsilon_{\rm sym}=0)\\
    &=& (\sigma_{xx}^{-1}(\varepsilon_{\rm sym}) - \sigma_{xx}^{-1}(\varepsilon_{\rm sym}=0))/\sigma_{xx}^{-1}(\varepsilon_{\rm sym}=0) \\
    &=& -\frac{\chi_{\rm sym}\varepsilon_{\rm sym} }{\chi_{\rm sym} \varepsilon_{\rm sym} +1}.
  \end{eqnarray*}
    At this time, the elastoresistivity is expressed as
    \begin{eqnarray*}
    {\rm ER}_{\rm sym} &=& \lim_{\varepsilon_{\rm sym}\to 0} \frac{ \Delta \rho_{xx} (\varepsilon_{\rm sym})/\rho_{xx}(\varepsilon_{\rm sym}=0)}{\varepsilon_{\rm sym} } \\
    &=& -\lim_{\varepsilon_{\rm sym}\to 0} \frac{\chi_{\rm sym}}{\chi_{\rm sym} \varepsilon_{\rm sym} +1}\\
    &=&-\chi_{\rm sym}.
    \end{eqnarray*}
    
Thus, a very simple result is obtained: ER$_{\rm sym} = - \chi_{\rm sym}$. Within the rigid band approximation, ER$_{\rm sym}$ purely reflects the strain-induced changes in carrier density.
    
      \begin{figure}[t]
      \centering
      \includegraphics[angle=0,width=85mm]{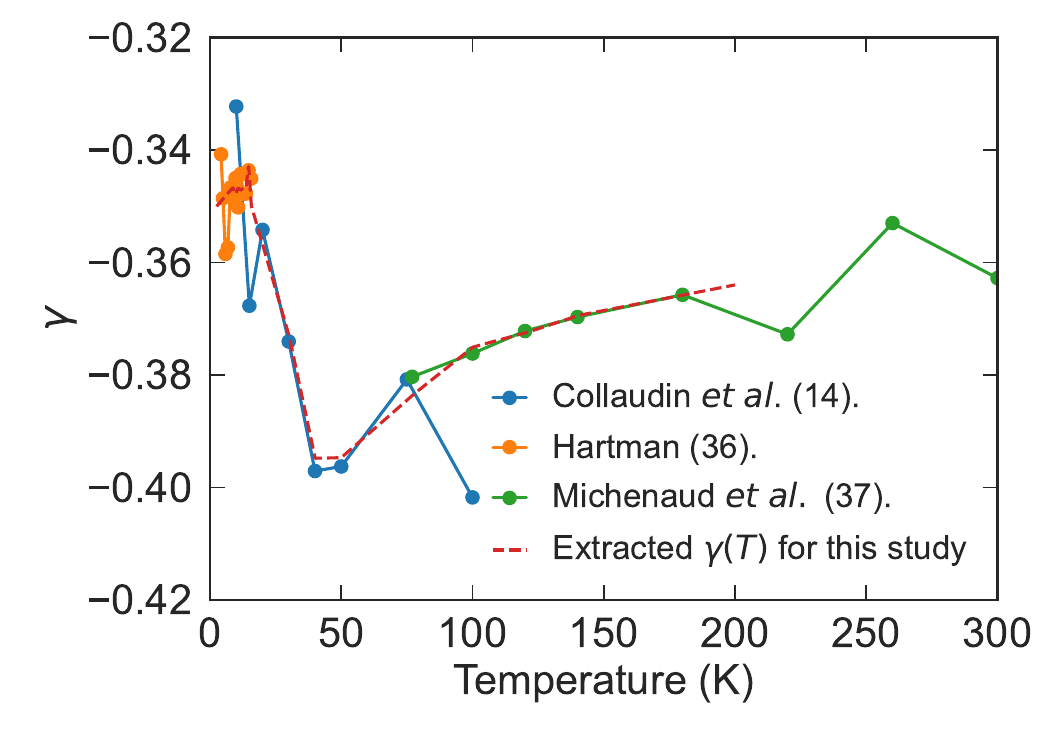}
      \caption{Temperature dependence of mobility tensor anisotropy $\gamma$ estimated from the previous studies\cite{PhysRevX.5.021022,PhysRev.181.1070,J_PMichenaud_1972}. In the main text, $\chi_{\rm anti}$ are evaluated by using the extracted temperature dependent $\gamma$ values indicated by the red dashed line.}
      \label{Gamma_Estimation}
      \end{figure} 
    
    Next, we discuss the case of antisymmetric strain $\frac{1}{2}(\varepsilon_{xx}- \varepsilon_{yy})$. The symmetry-breaking antisymmetric strain lifts the valley degeneracy and distinguishes one valley from the other two ones: $\Delta n_{{\rm e}1} = n \chi_{\rm anti} \varepsilon_{\rm anti}$ and $\Delta n_{{\rm e} 2/{\rm e} 3} = - n \chi_{\rm anti}\varepsilon_{\rm anti}/2$. This type of change in valley density is derived from the threefold symmetry of the system. This electron valley polarization does not change the total carrier number of electron valleys, and hence hole valley density is unaffected owing to the charge neutral conditions. In this case, the conductivity tensor under antisymmetric strain can be expressed as:
    \[
      \begin{split}
    \hat{\sigma} (\varepsilon_{\rm anti})
    = ne(&\hat{\mu}_{{\rm e}1} + \hat{\mu}_{{\rm e}2} + \hat{\mu}_{{\rm e}3}+ 3\hat{\nu} \\
    &+[\hat{\mu}_{{\rm e}1}-\hat{\mu}_{{\rm e}2}/2-\hat{\mu}_{{\rm e}3}/2]\chi_{\rm anti}\varepsilon_{\rm anti}).
    \end{split}
    \]
    Following the same procedure as the case of symmetric susceptibility, the elastoresistivity in antisymmetric symmetry is given as:
    \[
    \begin{split}
   {\rm ER}_{\rm anti} &=   \lim_{\varepsilon_{\rm anti}\to 0}\frac{ \Delta \rho_{xx} (\varepsilon_{\rm anti})/\rho_{xx}(\varepsilon_{\rm anti}=0)}{\varepsilon_{\rm anti} } \\
   &= -\frac{\mu^{{\rm e}1}_{xx}-\mu^{{\rm e}2}_{xx}/2-\mu^{{\rm e}3}_{xx}/2}{\mu^{{\rm e}1}_{xx} + \mu^{{\rm e}2}_{xx} + \mu^{{\rm e}3}_{xx}+ 3\nu_{xx}} \chi_{\rm anti}\\
   &= \gamma \chi_{\rm anti}.
      \end{split}
    \]
    
    Here, coefficient $\gamma$ represents the anisotropy of mobility, and thus ER$_{\rm anti}$ in the rigid band approximation is determined by the multiplications of strain-induced valley polarization with original valley anisotropy. Figure$\,$\ref{Gamma_Estimation} represents the temperature dependence of $\gamma$ evaluated by several previous studies for elucidating mobility tensors\cite{PhysRevX.5.021022,PhysRev.181.1070,J_PMichenaud_1972}. In the present experimental geometry, we measured the resistivity along abinary direction; therefore, the electron valley e1 has much larger mobility along this direction than those of electron valleys e2/e3, which results in a negative sign of $\gamma$. For the valley susceptibility analysis, we extracted the temperature dependence of $\gamma$ from the previous studies, as shown in Fig.$\,$\ref{Gamma_Estimation}.
   
\subsection*{3. Evaluations of valley susceptibilities through quantum oscillation measurements}
\begin{figure}
          \centering
          \includegraphics[angle=0,width=85mm]{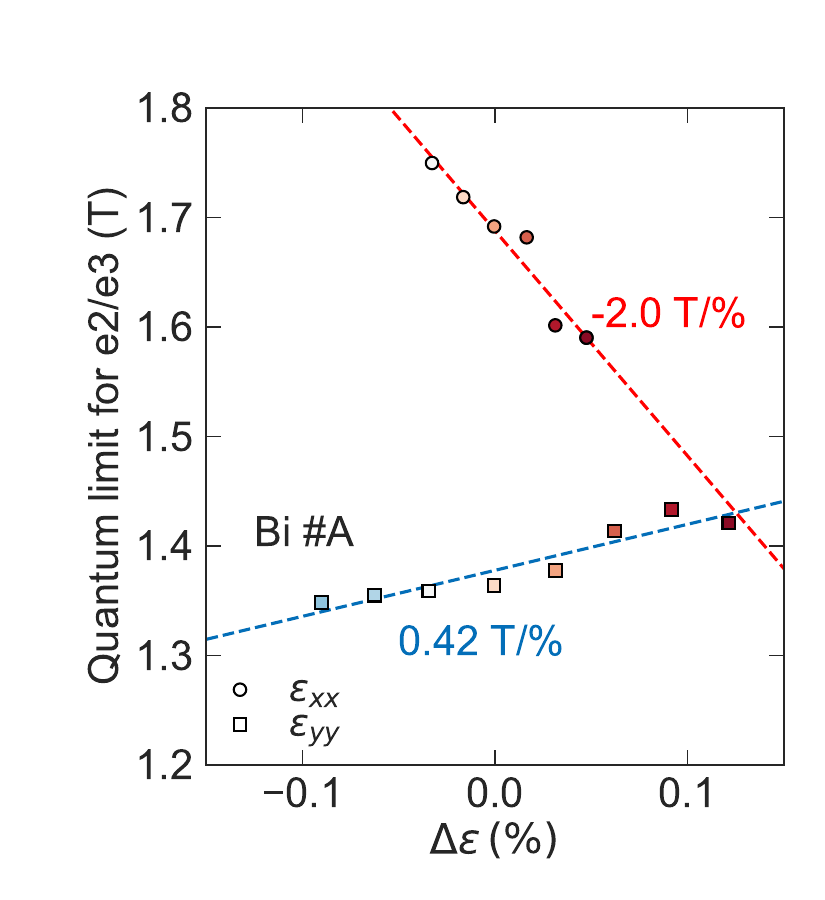}
          \caption{Strain dependent Shubnikov-de Haas measurements under field along binary for sample \#A. Quantum oscillation measurements are conducted at the base temperature of each cryostat we used: $T \sim 1.7$ K for the parallel geometry under $\varepsilon_{xx}$ and $T\sim 2$ K for the perpendicular geometry under $\varepsilon_{yy}$.}
          \label{Strain_dep_of_QL_e2e3xx_A}
          \end{figure}

  This simple carrier-based model allows us to evaluate the valley susceptibilities $\chi_{\rm sym}$ and $\chi_{\rm anti}$ from elastoresistance as demonstrated in the main text. Quantum oscillation provides a direct method for the evaluation of strain-induced changes in valley density. We have measured Shubnikov-de Haas oscillations under a magnetic field along binary direction for two samples \#A and \#B. Electron valleys e2/e3 reach the quantum limit at field of $\sim 1.5 \, {\rm T}$ along the binary direction. Positive strain $\varepsilon_{xx}$ in the parallel geometry shifts oscillation peaks toward the lower field side, while positive strain $\varepsilon_{yy}$ does them toward the opposite field side. This behavior is well reproduced in both samples; the results of sample \#A are shown in Fig.$\,$\ref{Strain_dep_of_QL_e2e3xx_A} and those of \#B are shown in the main text.

  In order to estimate the valley susceptibilities $\chi_{\rm sym}$ and $\chi_{\rm anti}$ from quantum oscillation results, we use the relation between the magnetic field at quantum limit $B_{\rm QL}$ and carrier density $n$ described as $B_{\rm QL} \propto n$. 
  Thus, the strain derivatives of $B_{\rm QL}$ for e2/e3 valleys give direct evaluations of changes in carrier densities of e2/e3 valleys against $\varepsilon_{xx}$ and $\varepsilon_{yy}$, respectively:
   \[
   \frac{1}{n_{{\rm e}2/{\rm e}3}(\varepsilon_{xx}=0)} \frac{{\rm d} n_{{\rm e}2/{\rm e}3}}{{\rm d} \varepsilon_{xx}} = \frac{1}{B_{\rm QL}^{e2/e3}(\varepsilon_{xx}=0)} \frac{{\rm d}B_{\rm QL}^{e2/e3}}{{\rm d}\varepsilon_{xx}},
   \]
   \[
      \frac{1}{n_{{\rm e}2/{\rm e}3}(\varepsilon_{yy}=0)} \frac{{\rm d} n_{{\rm e}2/{\rm e}3}}{{\rm d} \varepsilon_{yy}} = \frac{1}{B_{\rm QL}^{e2/e3}(\varepsilon_{yy}=0)} \frac{{\rm d}B_{\rm QL}^{e2/e3}}{{\rm d}\varepsilon_{yy}}.
   \]
   
      \begin{table}[t]
  \caption{ Approximate evaluations of valley susceptibilities in low temperatures by two different methods for sample \#A and \#B. $\chi^{\rm ER}$ is deduced from the value of elastoresistance at 2.5 K, and $\chi^{\rm QO}$ is evaluated from quantum oscillations at around 2 K.}
  \label{table:valley}
  \centering
  \begin{tabular}{|l||c|c||c|c|}
  \hline 
  Sample  & $\chi_{\rm sym}^{\rm ER}$ & $\chi_{\rm sym}^{\rm QO}$ &$\chi_{\rm anti}^{\rm ER}$&$\chi_{\rm anti}^{\rm QO}$\\
  \hline \hline
  \# A & -70& -100 & 85 & 250  \\
  \hline
  \# B & -95 & -110 & 190 &  280\\
  \hline
  \end{tabular}	
 \end{table}
   
   Here, we assume that the transverse strain direction is determined by the Poisson ratio of the platform $\nu_{\rm p}$, which determines the amount of induced symmetric and antisymmetric strains, respectively. In this case, the observed changes in valley density of e2/e3 valleys are described by valley susceptibilities as follows:
   \[
    \begin{split}
    & \frac{1}{n_{{\rm e}2/3}  (\varepsilon_{xx}=0)} \frac{{\rm d}  n_{{\rm e}2/{\rm e}3}}{{\rm d} \varepsilon_{xx}}  \\
   & =(1-\nu_{\rm p}) \chi_{\rm sym} \varepsilon_{\rm sym} /2 - (1+\nu_{\rm p})(\chi_{\rm anti}/2)\varepsilon_{\rm anti}/2,
   \end{split}
   \]

   \[
    \begin{split}
      &\frac{1}{n_{{\rm e}2/{\rm e}3}(\varepsilon_{yy}=0)} \frac{{\rm d} n_{{\rm e}2/{\rm e}3}}{{\rm d} \varepsilon_{yy}}\\
      &=   (1-\nu_{\rm p}) \chi_{\rm sym} \varepsilon_{\rm sym} /2 + (1+\nu_{\rm p})(\chi_{\rm anti}/2)\varepsilon_{\rm anti}/2.
    \end{split}
   \]
   By using these relations, the valley susceptibilities $\chi^{\rm QO}$ estimated from qunatum oscillation are obtained (results summarized in Table \ref{table:valley}). These values qualitatively agree with $\chi^{\rm ER}$ evaluated by the simple carrier model, including their signs and magnitudes, which suggests that our proposed simple carrier transport model successfully captures the essential nature of transport under stain in bismuth.

\subsection*{4. Valley susceptibility evaluated by deformation potentials}
The relationships between valley density and strain have been also discussed in magnetostriction study\cite{PhysRevB.26.2552}. Here, we developed a different method for evaluating the valley susceptibility as $\chi^{\rm MR}$
in accordance with the deformation potential-based framework introduced for describing previous magnetostriction results\cite{PhysRevB.26.2552}. Our analyzed $\chi^{\rm MR}$ gives consistent results with our elastoresistnace and quantum oscillation results. The detailed derivation of $\chi^{\rm MR}$ is described as follows.

  The magnetostrictive strain can be described by the field-induced change in the carrier densities\cite{PhysRevB.26.2552}. To describe the proportional relations between strain $\varepsilon_{ij}$ and energy shift of the band-edges $W_{\rm e,h}$ for electron and hole valleys, the deformation potential constants are introduced as below:
\[
W_{\rm e} = L_{ij} \varepsilon_{ij},
\]
\[
    W_{\rm h} = T_{ij} \varepsilon_{ij},
\]
  where the $L_{ij}$ are the components of the deformation potential tensor for the e1 electron valley and $T_{ij}$ are those for the hole valley. The values of these defomation potential constants determined by both experiment and theoretical calculations were reported in Ref.\cite{PhysRev.174.782,doi:10.1143/JPSJ.26.696}. According to Ref.\cite{PhysRevB.26.2552}, we can calculate the band shifts for electron valleys in general form:
\[
W_{\rm e1}  = L_{11} \varepsilon_{11} + L_{22} \varepsilon_{22} + L_{33} \varepsilon_{33} + 2 L_{23} \varepsilon_{23},
\]
\[
\begin{split}
    W_{\rm e2/e3} = \frac{1}{4}(L_{11}+3L_{22})\varepsilon_{11} + \frac{1}{4} (3 L_{11}+ L_{22}) \varepsilon_{22} + L_{33} \varepsilon_{33}\\
     \pm \frac{\sqrt{3}}{2} (L_{11}-L_{22} )\varepsilon_{12} \pm \sqrt{3} L_{23}\varepsilon_{13} -L_{23}\varepsilon_{23},
\end{split}
\]
where the upper/lower sign refers to e2/3 valley, respectively. Once we apply static stress along bianry or bisectix direction, 4 strain tensor components $\varepsilon_{11},\varepsilon_{22},\varepsilon_{33},\varepsilon_{23}$ are active\cite{10.1063/1.1735888}, and band shifts of electron valleys are given by:
\[
 W_{\rm e1} = L_{11} \varepsilon_{11} + L_{22} \varepsilon_{22} + L_{33}\varepsilon_{33} + 2L_{23} \varepsilon_{23},
\]
\[
\begin{split}
    W_{\rm e2/e3} = \frac{1}{4}(L_{11}+3L_{22}) \varepsilon_{11} + \frac{1}{4} (3L_{11} + L_{22}) \varepsilon_{22}\\
     +  L_{33} \varepsilon_{33} - L_{23} \varepsilon_{23}.
    \end{split}
\]
When stress is applied along the binary or bisectrix direcction just like in our study, there are no shear strain components $\varepsilon_{12}$ and $\varepsilon_{13}$, and thus there are no differences in band energy shift between e2 and e3 valleys. When longitudinal strain $\varepsilon$ along binary are induced, we can compute the energy shifts of bands:

\[
  W_{\rm e1}^{\sigma || {\rm binary}}  = L_{11} \varepsilon - L_{22} \nu_{12}\varepsilon - L_{33} \nu_{13} \varepsilon+ 2L_{23} \nu_{14} \varepsilon,
\]

\[
\begin{split}
  W_{\rm e2/e3}^{\sigma || {\rm binary}}  =\frac{1}{4}(L_{11}+3L_{22}) \varepsilon -\frac{1}{4} (3L_{11} + L_{22})  \nu_{12}\varepsilon \\
   - L_{33} \nu_{13} \varepsilon - L_{23} \nu_{14} \varepsilon.
 \end{split}
\]
Here, $\nu_{12},\nu_{13},\nu_{14}$ are the relevant poisson ratios of bimsuth, respectively. On the other hand, these band shifts can be described when the same amount of strain $\varepsilon$ is induced by the stress along bisectrix:

\[
  W_{\rm e1}^{\sigma || {\rm bisectrix}} = - L_{11} \nu_{12}  \varepsilon + L_{22} \varepsilon - L_{33} \nu_{13} \varepsilon -  2L_{23} \nu_{14} \varepsilon,
\]

\[
\begin{split}
  W_{\rm e2/e3}^{\sigma || {\rm bisectrix}}  = -\frac{1}{4}(L_{11}+3L_{22}) \nu_{12}  \varepsilon &+ \frac{1}{4} (3L_{11} + L_{22}) \varepsilon \\
   &- L_{33} \nu_{13} \varepsilon + L_{23} \nu_{14} \varepsilon.
  \end{split}
\]
To compare with our quantum oscillation results of e2/e3 valleys, we calculate the valley susceptibililities from $W_{\rm e2/e3}^{\sigma || {\rm binary}}$ and $W_{\rm e2/e3}^{\sigma || {\rm bisectrix}}$:
\[
\begin{split}
\chi_{\rm sym}^{\rm MR} &=- \frac{1}{\epsilon_e} \frac{{\rm d} W_{\rm e2/e3}^{\sigma || {\rm binary}}+W_{\rm e2/e3}^{\sigma || {\rm bisectrix}}}{{\rm d} \varepsilon} \\
 &= -\frac{1}{\epsilon_e} [ (L_{11}+ L_{22})(1-\nu_{12}) - 2 L_{33}\nu_{13}],
\end{split}
\]

\[
\begin{split}
  \chi_{\rm anti}^{\rm MR} &= -\frac{1}{\epsilon_e} \frac{{\rm d} W_{\rm e2/e3}^{\sigma || {\rm binary}}-W_{\rm e2/e3}^{\sigma || {\rm bisectrix}}}{{\rm d} \varepsilon}\\
   &= -\frac{1}{\epsilon_e} [\frac{-1}{2} (L_{11} - L_{22})(1+\nu_{12}) - 2 L_{23}\nu_{14}],
 \end{split}
\]
  where $\epsilon_{\rm e}$ represents the Fermi energy of the original electron valleys(typically $\epsilon_{\rm e} \sim \, 27 $ meV)\cite{PhysRevB.52.1566}. Poisson ratios $\nu_{12},\nu_{13},\nu_{14}$ can be caluclated from elastic constants $C_{ij}$.
 Combing deformation potential constants \cite{PhysRev.174.782} and elastic constants at 4 K \cite{Lichnowski_1976}, we can compute valley susceptibilities as $\chi_{\rm sym}^{\rm MR} \sim -155$ and $\chi_{\rm anti}^{\rm MR} \sim 280$, which are again consistent with our results shown in Table \ref{table:valley}. There facts further strenghten our conclusion on the strain-controlled valley density.

\subsection*{5. Consistency with DFT results}
As discussed in the main text, the strain-induced energy shifts of each valleys with rigid band nature are qualitatively consistent with our evaluated valley suseptibilities. In general, the GGA approximation is not enough to reproduce quantitative band structure such as band gap, as mentioned in the main text. Nevertheless, our results may quantitatively explain the results of $\chi_{\rm sym}$ and $\chi_{\rm anti}$. For example, antisymmetric strain $\chi_{\rm anti} \sim 0.5 \%$ is enough to induce complete valley polization emptying e1 or e2/e3 valleys, which is consistent with the experimental results $\chi_{\rm anti} > 200$.  On the other hand, it requires over $\varepsilon_{\rm sym} \sim 1 \%$ to induce metal-insulator transition. This difference is consistent with the fact that the $\chi_{\rm anti}$ exceeds $\chi_{\rm sym}$ by a factor of about 2.

%


\begin{thebibliography}{10}

\bibitem{SchaibleyNatRevMat2016}
J.~R. Schaibley, H.~Yu, G.~Clark, P.~Rivera, J.~S. Ross, K.~L. Seyler, W.~Yao,
  X.~Xu, Valleytronics in {2D} materials.
\newblock {\it Nature Reviews Materials\/} {\bf 1}, 16055 (2016).

\bibitem{PhysRevLett.97.186404}
O.~Gunawan, Y.~P. Shkolnikov, K.~Vakili, T.~Gokmen, E.~P. De~Poortere,
  M.~Shayegan, Valley susceptibility of an interacting two-dimensional electron
  system.
\newblock {\it Phys. Rev. Lett.\/} {\bf 97}, 186404 (2006).

\bibitem{IsbergNatMater2013}
J.~Isberg, M.~Gabrysch, J.~Hammersberg, S.~Majdi, K.~K. Kovi, D.~J. Twitchen,
  Generation, transport and detection of valley-polarized electrons in diamond.
\newblock {\it Nature Materials\/} {\bf 12}, 760--764 (2013).

\bibitem{ZengNatNanotech2012}
H.~Zeng, J.~Dai, W.~Yao, D.~Xiao, X.~Cui, Valley polarization in {MoS$_2$}
  monolayers by optical pumping.
\newblock {\it Nature Nanotechnology\/} {\bf 7}, 490--493 (2012).

\bibitem{MakNanotecqh2012}
K.~F. Mak, K.~He, J.~Shan, T.~F. Heinz, Control of valley polarization in
  monolayer {MoS$_2$} by optical helicity.
\newblock {\it Nature Nanotechnology\/} {\bf 7}, 494--498 (2012).

\bibitem{CaoNC2012}
T.~Cao, G.~Wang, W.~Han, H.~Ye, C.~Zhu, J.~Shi, Q.~Niu, P.~Tan, E.~Wang,
  B.~Liu, J.~Feng, Valley-selective circular dichroism of monolayer molybdenum
  disulphide.
\newblock {\it Nature Communications\/} {\bf 3}, 887 (2012).

\bibitem{ZhuNatPhys2012}
Z.~Zhu, A.~Collaudin, B.~Fauqu{\'e}, W.~Kang, K.~Behnia, Field-induced
  polarization of dirac valleys in bismuth.
\newblock {\it Nature Physics\/} {\bf 8}, 89--94 (2012).

\bibitem{KulcherNatMater2014}
R.~K{\"u}chler, L.~Steinke, R.~Daou, M.~Brando, K.~Behnia, F.~Steglich,
  Thermodynamic evidence for valley-dependent density of states in bulk
  bismuth.
\newblock {\it Nature Materials\/} {\bf 13}, 461--465 (2014).

\bibitem{JPIssi}
J.-P. Issi, Low temperature transport properties of the group {V} semimetals.
\newblock {\it Australian Journal of Physics\/} {\bf 32}, 585--628 (1979).

\bibitem{doi:10.7566/JPSJ.84.012001}
Y.~Fuseya, M.~Ogata, H.~Fukuyama, Transport properties and diamagnetism of
  dirac electrons in bismuth.
\newblock {\it Journal of the Physical Society of Japan\/} {\bf 84}, 012001
  (2015).

\bibitem{Zhu_2018}
Z.~Zhu, B.~Fauqu{\'e}, K.~Behnia, Y.~Fuseya, Magnetoresistance and valley
  degree of freedom in bulk bismuth.
\newblock {\it Journal of Physics: Condensed Matter\/} {\bf 30}, 313001 (2018).

\bibitem{ZhuNatCommun2017}
Z.~Zhu, J.~Wang, H.~Zuo, B.~Fauqu{\'e}, R.~D. McDonald, Y.~Fuseya, K.~Behnia,
  Emptying dirac valleys in bismuth using high magnetic fields.
\newblock {\it Nature Communications\/} {\bf 8}, 15297 (2017).

\bibitem{IwasaSciRep2019}
A.~Iwasa, A.~Kondo, S.~Kawachi, K.~Akiba, Y.~Nakanishi, M.~Yoshizawa,
  M.~Tokunaga, K.~Kindo, Thermodynamic evidence of magnetic-field-induced
  complete valley polarization in bismuth.
\newblock {\it Scientific Reports\/} {\bf 9}, 1672 (2019).

\bibitem{PhysRevX.5.021022}
A.~Collaudin, B.~Fauqu\'e, Y.~Fuseya, W.~Kang, K.~Behnia, Angle dependence of
  the orbital magnetoresistance in bismuth.
\newblock {\it Phys. Rev. X\/} {\bf 5}, 021022 (2015).

\bibitem{BrandtSovPhys}
N.~B. Brandt, V.~A. Kul'bachinkii, N.~Y. Minina, V.~D. Shirokikh, Change of the
  band structure and electronic phase transitions in {Bi} and
  {Bi$_{1-x}$Sb$_{x}$} alloys under uniaxial tension strains.
\newblock {\it Sov. Phys. JETP\/} {\bf 51}, 562 (1980).

\bibitem{ZhouNatMat}
S.~Y. Zhou, G.~H. Gweon, A.~V. Fedorov, P.~N. First, W.~A. de~Heer, D.~H. Lee,
  F.~Guinea, A.~H. Castro~Neto, A.~Lanzara, Substrate-induced bandgap opening
  in epitaxial graphene.
\newblock {\it Nature Materials\/} {\bf 6}, 770--775 (2007).

\bibitem{JurgenNatPhoto}
J.~Michel, J.~Liu, L.~C. Kimerling, High-performance {Ge-on-Si} photodetectors.
\newblock {\it Nature Photonics\/} {\bf 4}, 527--534 (2010).

\bibitem{PhysRevLett.105.136805}
K.~F. Mak, C.~Lee, J.~Hone, J.~Shan, T.~F. Heinz, Atomically thin
  {$\mathrm{MoS}_{2}$}: A new direct-gap semiconductor.
\newblock {\it Phys. Rev. Lett.\/} {\bf 105}, 136805 (2010).

\bibitem{HOFMANN2006191}
P.~Hofmann, The surfaces of bismuth: Structural and electronic properties.
\newblock {\it Progress in Surface Science\/} {\bf 81}, 191--245 (2006).

\bibitem{10.1063/5.0008829}
J.~Park, J.~M. Bartlett, H.~M.~L. Noad, A.~L. Stern, M.~E. Barber,
  M.~K{\"{o}}nig, S.~Hosoi, T.~Shibauchi, A.~P. Mackenzie, A.~Steppke, C.~W.
  Hicks, {Rigid platform for applying large tunable strains to mechanically
  delicate samples}.
\newblock {\it Review of Scientific Instruments\/} {\bf 91}, 083902 (2020).

\bibitem{10.1063/1.4881611}
C.~W. Hicks, M.~E. Barber, S.~D. Edkins, D.~O. Brodsky, A.~P. Mackenzie,
  {Piezoelectric-based apparatus for strain tuning}.
\newblock {\it Review of Scientific Instruments\/} {\bf 85}, 065003 (2014).

\bibitem{PhysRevB.88.085113}
H.-H. Kuo, M.~C. Shapiro, S.~C. Riggs, I.~R. Fisher, Measurement of the
  elastoresistivity coefficients of the underdoped iron arsenide
  {Ba(Fe${}_{0.975}$Co${}_{0.025}$)${}_{2}$As${}_{2}$}.
\newblock {\it Phys. Rev. B\/} {\bf 88}, 085113 (2013).

\bibitem{PhysRevB.44.7888}
G.~te~Velde, E.~J. Baerends, Precise density-functional method for periodic
  structures.
\newblock {\it Phys. Rev. B\/} {\bf 44}, 7888--7903 (1991).

\bibitem{BAND}
{\it {BAND} 2021.1\/} (2021). Https://www.scm.com/.

\bibitem{Liu1995}
Y.~Liu, R.~E. Allen, Electronic structure of the semimetals {B}i and {S}b.
\newblock {\it Phys. Rev. B\/} {\bf 52}, 1566 (1995).

\bibitem{PhysRevB.91.125129}
I.~Aguilera, C.~Friedrich, S.~Bl\"ugel, Electronic phase transitions of bismuth
  under strain from relativistic self-consistent {$GW$} calculations.
\newblock {\it Phys. Rev. B\/} {\bf 91}, 125129 (2015).

\bibitem{Koike_1966}
R.~Koike, H.~Kurokawa, Elastoresistance effects in evaporated bismuth films.
\newblock {\it Japanese Journal of Applied Physics\/} {\bf 5}, 503 (1966).

\bibitem{RevPhysAppl1979}
M.~Saleh, J.~Buxo, G.~Dorville, G.~Sarrabayrouse, Electrical and
  elastoresistance properties of evaporated thin films of bismuth.
\newblock {\it Rev. Phys. Appl.\/} {\bf 14}, 405--413 (1979).

\bibitem{PhysRev.135.A708}
A.~L. Jain, R.~Jaggi, Piezo-galvanomagnetic effects in bismuth.
\newblock {\it Phys. Rev.\/} {\bf 135}, A708--A710 (1964).

\bibitem{PhysRev.149.485}
R.~T. Bate, W.~E. Drobish, N.~G. Einspruch, Piezoresistivity of bismuth.
\newblock {\it Phys. Rev.\/} {\bf 149}, 485--489 (1966).

\bibitem{PRSL_Kapitza}
P.~Kapitza, The study of the magnetic properties of matter in strong magnetic
  fields. part{V}.-experiments on magnetostriction in dia- and para-magnetic
  substances.
\newblock {\it Proc. R. Soc. Lond. A\/} {\bf 135}, 568--600 (1932).

\bibitem{PRSL_Shoenberg}
D.~Shoenberg, The magnetostriction of bismuth single crystals.
\newblock {\it Proc. R. Soc. Lond. A\/} {\bf 150}, 619--637 (1932).

\bibitem{PhysRevLett.11.331}
B.~A. Green, B.~S. Chandrasekhar, Observation of oscillatory magnetostriction
  in bismuth at 4.2\ifmmode^\circ\else\textdegree\fi{}k.
\newblock {\it Phys. Rev. Lett.\/} {\bf 11}, 331--332 (1963).

\bibitem{JPMichenaud_1981}
J.~P. Michenaud, J.~Heremans, J.~Boxus, C.~Haumont, Longitudinal
  magnetostriction of bismuth above the last quantum limit.
\newblock {\it Journal of Physics C: Solid State Physics\/} {\bf 14}, L13
  (1981).

\bibitem{PhysRevB.26.2552}
J.~P. Michenaud, J.~Heremans, M.~Shayegan, C.~Haumont, Magnetostriction of
  bismuth in quantizing magnetic fields.
\newblock {\it Phys. Rev. B\/} {\bf 26}, 2552--2559 (1982).

\bibitem{PhysRev.181.1070}
R.~Hartman, Temperature dependence of the low-field galvanomagnetic
  coefficients of bismuth.
\newblock {\it Phys. Rev.\/} {\bf 181}, 1070--1086 (1969).

\bibitem{J_PMichenaud_1972}
J.~P. Michenaud, J.~P. Issi, Electron and hole transport in bismuth.
\newblock {\it Journal of Physics C: Solid State Physics\/} {\bf 5}, 3061
  (1972).

\bibitem{doi:10.1073/pnas.1910695116}
N.~H. Jo, L.-L. Wang, P.~P. Orth, S.~L. Bud{'}ko, P.~C. Canfield,
  Magnetoelastoresistance in {WTe$_2$}: Exploring electronic structure and
  extremely large magnetoresistance under strain.
\newblock {\it Proceedings of the National Academy of Sciences\/} {\bf 116},
  25524--25529 (2019).

\bibitem{doi:10.1126/science.1221713}
J.-H. Chu, H.-H. Kuo, J.~G. Analytis, I.~R. Fisher, Divergent nematic
  susceptibility in an iron arsenide superconductor.
\newblock {\it Science\/} {\bf 337}, 710--712 (2012).

\bibitem{doi:10.1126/science.aab0103}
H.-H. Kuo, J.-H. Chu, J.~C. Palmstrom, S.~A. Kivelson, I.~R. Fisher, Ubiquitous
  signatures of nematic quantum criticality in optimally doped {F}e-based
  superconductors.
\newblock {\it Science\/} {\bf 352}, 958--962 (2016).

\bibitem{doi:10.1073/pnas.1605806113}
S.~Hosoi, K.~Matsuura, K.~Ishida, H.~Wang, Y.~Mizukami, T.~Watashige,
  S.~Kasahara, Y.~Matsuda, T.~Shibauchi, Nematic quantum critical point without
  magnetism in {FeSe$_{1-x}$S$_x$} superconductors.
\newblock {\it Proceedings of the National Academy of Sciences\/} {\bf 113},
  8139--8143 (2016).

\bibitem{doi:10.1073/pnas.2110501119}
K.~Ishida, Y.~Onishi, M.~Tsujii, K.~Mukasa, M.~Qiu, M.~Saito, Y.~Sugimura,
  K.~Matsuura, Y.~Mizukami, K.~Hashimoto, T.~Shibauchi, Pure nematic quantum
  critical point accompanied by a superconducting dome.
\newblock {\it Proceedings of the National Academy of Sciences\/} {\bf 119},
  e2110501119 (2022).

\bibitem{PhysRevB.84.115137}
Z.~Zhu, B.~Fauqu\'e, Y.~Fuseya, K.~Behnia, Angle-resolved landau spectrum of
  electrons and holes in bismuth.
\newblock {\it Phys. Rev. B\/} {\bf 84}, 115137 (2011).

\bibitem{PhysRev.174.782}
K.~Walther, Anisotropy of magnetoacoustic attenuation and deformation potential
  in bismuth.
\newblock {\it Phys. Rev.\/} {\bf 174}, 782--790 (1968).

\bibitem{LittleArielleNatMater2020}
A.~Little, C.~Lee, C.~John, S.~Doyle, E.~Maniv, N.~L. Nair, W.~Chen, D.~Rees,
  J.~W.~F. Venderbos, R.~M. Fernandes, J.~G. Analytis, J.~Orenstein,
  Three-state nematicity in the triangular lattice antiferromagnet
  {Fe$_{1/3}$NbS$_2$}.
\newblock {\it Nature Materials\/} {\bf 19}, 1062--1067 (2020).

\bibitem{LinpengNature2022}
L.~Nie, K.~Sun, W.~Ma, D.~Song, L.~Zheng, Z.~Liang, P.~Wu, F.~Yu, J.~Li,
  M.~Shan, D.~Zhao, S.~Li, B.~Kang, Z.~Wu, Y.~Zhou, K.~Liu, Z.~Xiang, J.~Ying,
  Z.~Wang, T.~Wu, X.~Chen, Charge-density-wave-driven electronic nematicity in
  a kagome superconductor.
\newblock {\it Nature\/} {\bf 604}, 59--64 (2022).

\bibitem{ChangwooNC2020}
C.-w. Cho, J.~Shen, J.~Lyu, O.~Atanov, Q.~Chen, S.~H. Lee, Y.~S. Hor, D.~J.
  Gawryluk, E.~Pomjakushina, M.~Bartkowiak, M.~Hecker, J.~Schmalian, R.~Lortz,
  {$Z_3$}-vestigial nematic order due to superconducting fluctuations in the
  doped topological insulators {Nb$_x$Bi$_2$Se$_3$ and Cu$_x$Bi$_2$Se$_3$}.
\newblock {\it Nature Communications\/} {\bf 11}, 3056 (2020).

\bibitem{doi:10.1126/science.aag1715}
B.~E. Feldman, M.~T. Randeria, A.~Gyenis, F.~Wu, H.~Ji, R.~J. Cava, A.~H.
  MacDonald, A.~Yazdani, Observation of a nematic quantum hall liquid on the
  surface of bismuth.
\newblock {\it Science\/} {\bf 354}, 316--321 (2016).

\bibitem{KangNC2022}
W.~Kang, F.~Spathelf, B.~Fauqu{\'e}, Y.~Fuseya, K.~Behnia, Boundary conductance
  in macroscopic bismuth crystals.
\newblock {\it Nature Communications\/} {\bf 13}, 189 (2022).

\bibitem{JEAubrey_1971}
J.~E. Aubrey, Magnetoconductivity tensor for semimetals.
\newblock {\it Journal of Physics F: Metal Physics\/} {\bf 1}, 493 (1971).

\bibitem{doi:10.1143/JPSJ.26.696}
S.-i. Katsuki, Calculation of deformation potentials in bismuth.
\newblock {\it Journal of the Physical Society of Japan\/} {\bf 26}, 696--700
  (1969).

\bibitem{10.1063/1.1735888}
Y.~Eckstein, A.~W. Lawson, D.~H. Reneker, {Elastic Constants of Bismuth}.
\newblock {\it Journal of Applied Physics\/} {\bf 31}, 1534--1538 (2004).

\bibitem{PhysRevB.52.1566}
Y.~Liu, R.~E. Allen, Electronic structure of the semimetals bi and sb.
\newblock {\it Phys. Rev. B\/} {\bf 52}, 1566--1577 (1995).

\bibitem{Lichnowski_1976}
A.~J. Lichnowski, G.~A. Saunders, The elastic constants of bismuth-antimony
  alloy single crystals.
\newblock {\it Journal of Physics C: Solid State Physics\/} {\bf 9}, 927
  (1976).

\end{thebibliography}

\end{document}